# Investigation of high pressure capacitively coupled plasmas produced by electrons energized in DC sheath at powered electrode

**Anuravi Sharma[1], Ramesh Narayanan[1], Arti Rawat[2], Ashish Ganguli[1]**

[1]Department of Energy Science & Engineering, Indian Institute of Technology Delhi, New Delhi-110016, India

[2]Applied Materials Dublin Ireland

***Author to whom correspondence should be addressed:** rams@dese.iitd.ac.in

**Abstract:** A 13.56 MHz, capacitively coupled plasma is investigated experimentally to determine the power absorption mechanism across a wide pressure range (≈ 5–600 mTorr) at ≈ 10 W. Axial profiles of plasma parameters are measured along with $V_{DC}$, the DC self-bias voltage on the powered electrode (PE), from which the DC sheath voltage drop, $V_s$ is determined. Axial profiles of electron ohmic power absorption reveal that power deposition is highest in low-density regions and lowest in high-density regions, indicating that plasma formation is not driven by Ohmic heating. Probability arguments were correlated with locations of the density peaks to determine the ionization mean free paths ($\lambda_{iz}$) at each pressure. Electron acceleration and average electron velocity acquired in the DC sheath voltage drop at PE were also calculated to determine $\lambda_{iz}$ independently for comparing with those calculated from the density profiles. The agreement is good for all pressures, barring the lowest pressure (≈ 5 mTorr) for which there is significant deviation. The electron sheath transit times are a small fraction of the RF period, implying that the accelerated electrons experience the RF field as instantaneous "spot values" superimposed on $V_s$, the DC sheath drop. The negative self-bias on PE renders the RF swing asymmetric making it negative for most of the cycle. Since the RF accelerates electrons when it is negative and takes energy from them when it is positive, net power is transferred to the electrons during the course of steady-state measurements, *averaged over many RF cycles*. Except at ≈ 5 mTorr where stochastic heating dominates, the RF field at higher pressures exhibits a novel role, not hitherto reported.

## 1 Introduction

Capacitively Coupled Plasmas (CCPs) are widely employed in plasma-based processes such as etching, sputtering, thin-film deposition, and surface cleaning, forming the foundation for various semiconductor and biomedical applications. Despite a reasonably well-established understanding of their fundamental operation, there is also a need to meet the demands of the industry and so one needs to optimize the discharge systems for better technical performance. The growing performance demands necessitate continuous, in-depth investigations and comprehensive characterizations of the discharge. In this regard, a critical challenge lies in the complex nature of RF power coupling within these discharges, which significantly influences their generation, sustainment, and control [1-7]. The absorption of RF power by the plasma facilitates electron heating, a key mechanism necessary for effective ionization and sustained plasma generation. In order to achieve efficient control over the particle and energy fluxes striking the electrodes, a detailed analysis of electron power absorption and plasma parameters is also necessary.

The spatial variations exhibited by the various mechanisms of power absorption by electrons and their roles in different pressure regimes play crucial role in governing the overall plasma dynamics. For this concern, numerous studies have extensively investigated the role of electron heating mechanisms (*primarily Ohmic and stochastic heating*) in CCPs using Particle-in-Cell/Monte Carlo Collision (PIC/MCC) simulations and theoretical models. Surendra and Dalvie were the first to analyse self-consistently power absorption by electrons in an electropositive, single-frequency, low-pressure CCP using a combined Boltzmann equation and PIC simulation framework [8]. Their study revealed that the electron pressure-gradient term—previously neglected in simplified descriptions of *ohmic heating*—plays a significant role in driving *collisionless* (stochastic) power absorption. Turner *et al*. have extensively explored the theoretical understanding of stochastic heating in low pressure CCPs and presented another mechanism of electron heating *viz*. 'pressure heating' that arises from the spatial gradients in the electron pressure, which become significant in low-pressure and strongly oscillating plasmas such as capacitively coupled plasmas [9-11]. Kaganovich *et al* gave a theoretical understanding of stochastic (collisionless) heating in low pressure bounded plasmas [12]. G.Y. Park et al also studied the collision less heating in CCPs at low pressures using PIC simulations focussing on the interaction of RF electric field with the electron motion [13]. Various studies revealed an enhanced heating regime due to synchronized motion between electron sheath edge oscillations and the motion of the electrons reflected by the sheath edge [14,15]. T. Mussenbrock *et al* investigated self-excited resonance effects inside plasma that can enhance the ohmic and stochastic power absorption. A recent study by Vass *et al*. based on PIC/Monte Carlo simulations revealed ohmic power absorption as the dominant mechanism at low pressures [16]. These simulations and theoretical studies provide detailed insights into the dominance of electron power absorption mechanisms under various discharge conditions. However, most of these studies primarily rely on simulations or theoretical models, and experimentally validated understanding of electron-heating in different regimes has remained scant. Although stochastic power absorption has been widely studied and established as a dominant mechanism at low pressures based on theoretical and numerical investigations [17-19], experimental investigations of power absorption in moderately high to high pressure regimes is still limited. Recently, Rawat *et al*. presented experimental studies on power absorption at high pressures [20]. *Using a newly developed diagnostic technique, the J·E probe, direct measurements of the spatial distribution of power absorption / deposition within the discharge were reported that gave indication of strong RF power absorption near the powered electrode even at high pressures* [21].

This paper attempts to present evidence for plasma generation at high pressures by electrons energized through DC sheath drop at powered electrode (in CCPs). To this end, two measurements were performed. (a) Detailed profiling of the plasma parameters along the system length using a compensated Langmuir probe (LP) was undertaken for a wide range of pressures ≈ 5 – 600 mTorr. (b) The DC self-bias voltage, $V_{DC}$ induced on the powered electrode was also measured at the different pressures. The latter data can be used to determine the DC voltage drop (= $V_s$) across the steady state sheath at the powered electrode (PE). The collected data allow one to generate axial profiles of $P_{ohmic}$, the Ohmic power absorbed by the bulk electrons. These profiles show clearly that $P_{ohmic}$ peaks in regions of low plasma density (or low conductivity) and is lowest in regions of high plasma density (or high conductivity), indicating clearly that Ohmic power absorption is not responsible for plasma generation in the experiments. To investigate the plasma formation mechanism in this study the following procedure was adopted. Firstly, using simple probability arguments and the locations of the plasma density peaks with respective to the powered electrode, the ionization mean free paths at each pressure were computed. Next, a steady state model of electron acceleration in the DC voltage drop (= $V_s$) inside the sheath at the powered electrode was used to determine the average directed velocity of the electrons within the sheath from which the ionization mean free paths were determined. Comparison of the mean free paths calculated from the two methods show very good agreement for most pressures (except the lowest, for which there was significant discrepancy). To clarify the role of the RF in the present experiments, the transit times of electrons through the sheath were determined. It turns out that these transit times are a small fraction of the RF period, implying that the accelerated electrons experience the RF field as instantaneous "spot values" superimposed on $V_s$, the DC sheath drop. The negative self-bias on PE renders the RF swing asymmetric making it negative for most of the cycle and since the RF accelerates electrons when it is negative and takes energy from them when it is positive, net power is transferred to the electrons during the course of steady-state measurements, *averaged over many RF cycles*.

Section 2 gives a summary of the relevant formulas used in this work [22]. Section 3 describes the experimental setup along with the plasma diagnostic techniques employed in this study. Section 4 presents the experimental results and Sections 5 analyses the results to develop a framework for understanding the plasma generation mechanism. Section 6 uses this framework to uncover the role of RF in the present experiments. Section 7 is the concluding section and summarises the results of the paper in the CCP.

## 2 Relevant formulas used

The *real* ohmic power $P_{ohmic}$ absorbed by the electrons of the bulk plasma in the system can be expressed as,

$$P_{ohmic} = \frac{1}{2} real\,(\boldsymbol{E}.\boldsymbol{J}^*) = \frac{1}{2} J^2 Real\left(\frac{1}{j\omega\epsilon_o\kappa_p}\right) \qquad \text{.....(1)}$$

where,

$$\boldsymbol{E} = \frac{\boldsymbol{J}}{j\omega\epsilon_o\kappa_p}, \qquad \kappa_p = 1 - \frac{\omega_{pe}^2}{\omega(\omega - j\nu_m)} \qquad \text{.....(2)}$$

Here $\boldsymbol{E}$ and $\boldsymbol{J}$ are the complex amplitudes of the RF electric field and current, respectively along the system axis. $\boldsymbol{J}$ is the total current and includes the conduction and displacement currents. $J^2 = \boldsymbol{J}.\boldsymbol{J}^*$ is the *absolute squared magnitude* of the complex amplitude $\boldsymbol{J}$ and $J = |\boldsymbol{J}|$, is the magnitude of $\boldsymbol{J}$. $\epsilon_o$ is the free space permittivity and $\kappa_p$ is the complex plasma dielectric function; $\nu_m$ is the momentum transfer frequency for elastic, electron – neutral collisions at electron temperature $T_e$. $\omega_{pe}$ is the plasma frequency (radians) and $\omega$ is the radian frequency of the applied RF.

To determine $J$, one resorts to the equations of the *inhomogeneous model* discussed in detail in Ref. 22 [Chapter 11, Section 11.2]. It is given as

$$J^2 = 2.88 e\epsilon_o\omega^2\sqrt{T_{es}V_{rf}}\,n_s \qquad \text{.....(3)}$$

In Eq. (3), $n_s$ and $T_{es}$ are the electron density and electron temperature *at the edge of the steady state sheath at the powered electrode* (PE). In the experiments these are obtained from Langmuir probe (LP) measurements at a location as close to the PE as possible. $V_{rf}$ in Eq. (3) is the *magnitude* of the RF voltage amplitude at the PE $(V_{rf} = |\boldsymbol{V}_{rf}|)$. To determine $V_{rf}$ one needs to go back to the inhomogeneous model to write,

$$V_s = 0.78V_{rf} \qquad \ldots..(4)$$

In Eq. (4), $V_s$ is the *DC voltage drop across steady state sheath at the PE* and has to be obtained indirectly from the measurements. It is given as,

$$V_s = V_{ps} - V_{DC} \qquad \ldots..(5)$$

Here $V_{ps}$ is the plasma potential at the steady state sheath edge at the PE (obtained from LP measurements as mentioned above) and $V_{DC}$ is the DC, self-bias voltage that develops on the PE. As stated earlier, in the present work, $V_{DC}$ was obtained by direct measurement. For later calculations, the width $s$ of the steady state sheath at the PE would be required. Once again, turning to the inhomogeneous model [22] one may write

$$s = \sqrt{\frac{1.16\epsilon_o}{en_s}\left(\frac{V_s{}^3}{T_{es}}\right)^{1/2}} \qquad \ldots..(6)$$

## 3 Experimental Setup

### 3.1 Capacitively Coupled plasma System

Experiments were conducted in an asymmetric RF discharge system featuring a parallel-plate electrode configuration and argon as the working gas. The setup included two stainless steel disc electrodes: a powered electrode (PE) with a diameter of 15 cm and a slightly larger grounded electrode (GE) measuring 15.7 cm [23]. The electrodes were positioned, $d$ = 7.5 cm apart, inside a grounded stainless-steel chamber. The grounded chamber along with the grounded electrode, comprised the surface area of the ground for the RF plasma discharge system. Hence this, introduced an asymmetry in the system due to differing RF powered and grounded electrode surface areas. The RF power was delivered from a 13.56 MHz RF generator (SEREN R1001) equipped with an automatic matching network controller (SEREN *Model MC2*) enabling efficient RF coupling with minimal losses. The matching network has a series or tuning capacitor that also blocks the DC that appears on the PE as a self-bias voltage. The chamber was evacuated to a base pressure ≈ $1 \times 10^{-5}$ Torr, and experiments were conducted over a wide range of argon gas pressures, from ≈ 5 to ≈ 600 mTorr and an RF power of ≈ 10W. A RF-compensated Langmuir Probe (CLP) [23] inserted from the bottom port of the chamber was used to measure plasma parameters spatially along the axial length of the discharge. Figure 1 shows a detailed schematic of Capacitively Coupled Plasma (CCP) system with the LP mounted on to the bottom port of the vacuum chamber.

### 3.2 Diagnostics used

To investigate the axial variation of power absorption mechanisms in the discharge, two diagnostic approaches were employed. The compensated Langmuir probe (CLP) shown in Figure 2(a) was utilized to measure the axial profiles of basic plasma parameters, such as plasma density ($n_e$), electron temperature ($T_e$) and plasma potential ($V_p$), which were used as essential inputs for the determination of the various quantities. A 10x voltage probe was employed to monitor the DC self-bias voltage that developed on the powered electrode including the drop across sheath at the PE. These aspects are discussed in the next section.

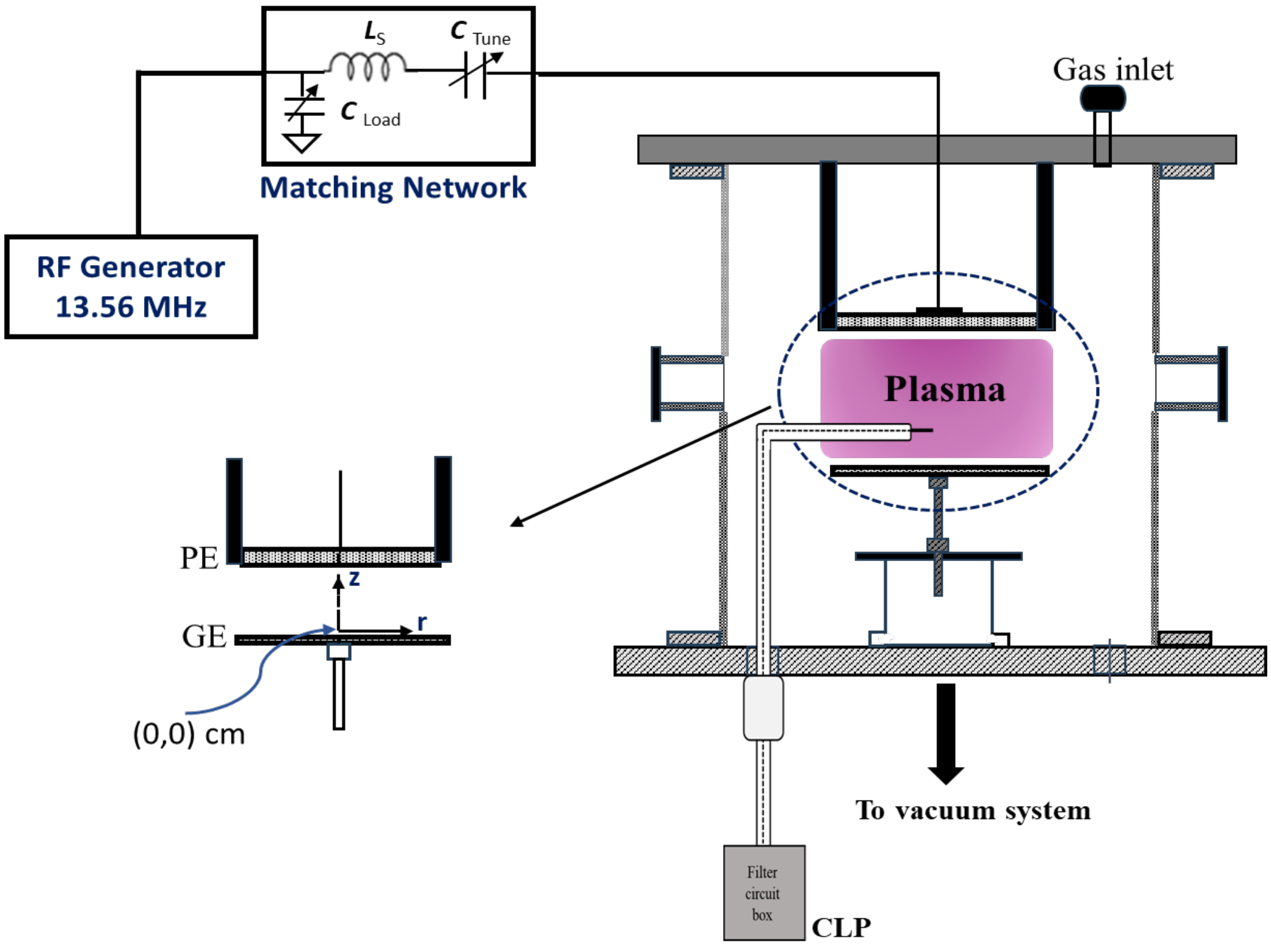

**Figure 1.** Schematic of Capacitively Coupled plasma Discharge System (CCP) with installed Compensated Langmuir probe.

### 3.2.1 Compensated Langmuir Probe (CLP)

It is a known fact that in RF plasmas, the rapid oscillations in plasma potential, typically at 13.56 MHz or other RF harmonic frequencies, can lead to distortion of the LP current-voltage (*I-V*) characteristics. So, a compensated LP was used to accurately measure plasma parameters such as plasma density ($n_e$), electron temperature ($T_e$) and plasma potential ($V_p$). A compensated Langmuir probe comprising a three-stage RF filter was employed in this work, which was specifically designed, fabricated, and tuned to attenuate the frequencies, 13.56 MHz (fundamental, *f*), 27.12 MHz (second harmonic, *2f*), and 40.68 MHz (third harmonic, *3f*), following the methodology outlined in [24]. The probe structure and its tuning methodology ensures that the probe remains tuned in the presence of plasma up to the 3rd harmonic, while ensuring that the total filter impedance ($Z_F$) at each harmonic frequency (up to the 3rd) is much greater than the sheath impedance, i.e., $|Z_F| \gg |Z'_{sh}|$. A detailed analysis of the probe with its equivalent circuit has been explained in Ref. 23.

Figure 2(a) shows the schematic of a Compensated Langmuir Probe (CLP) affixed to its filter-box. The filter response was measured using an impedance analyzer (Model No: Agilent 4294A). Figure 2(b) displays the filter response at the three harmonics. It can be seen that $|Z|$ peaks sharply at the first two harmonics giving impedances (≈ 20.5 and ≈1.25 kΩ respectively), although its response at the third is rather weak. Nonetheless, the filter ensures that the *I–V* characteristics of the LP are free from distortion due to the first two harmonics, which are the strongest. A picture of the plasma along with the CLP inserted into the plasma shown in figure 2(c). The probe assembly is inserted from an off-axis port on the end plate of the chamber that allows translation along the axis (z-axis) between the powered and grounded electrodes. Additionally, it can be rotated about its shifted axis, which allows the probe tip to sweep the plasma in an arc. The angle swept allows one to determine the radial location of the tip. Thus, one can obtain both axial and radial profiles of the plasma parameters using the probe.

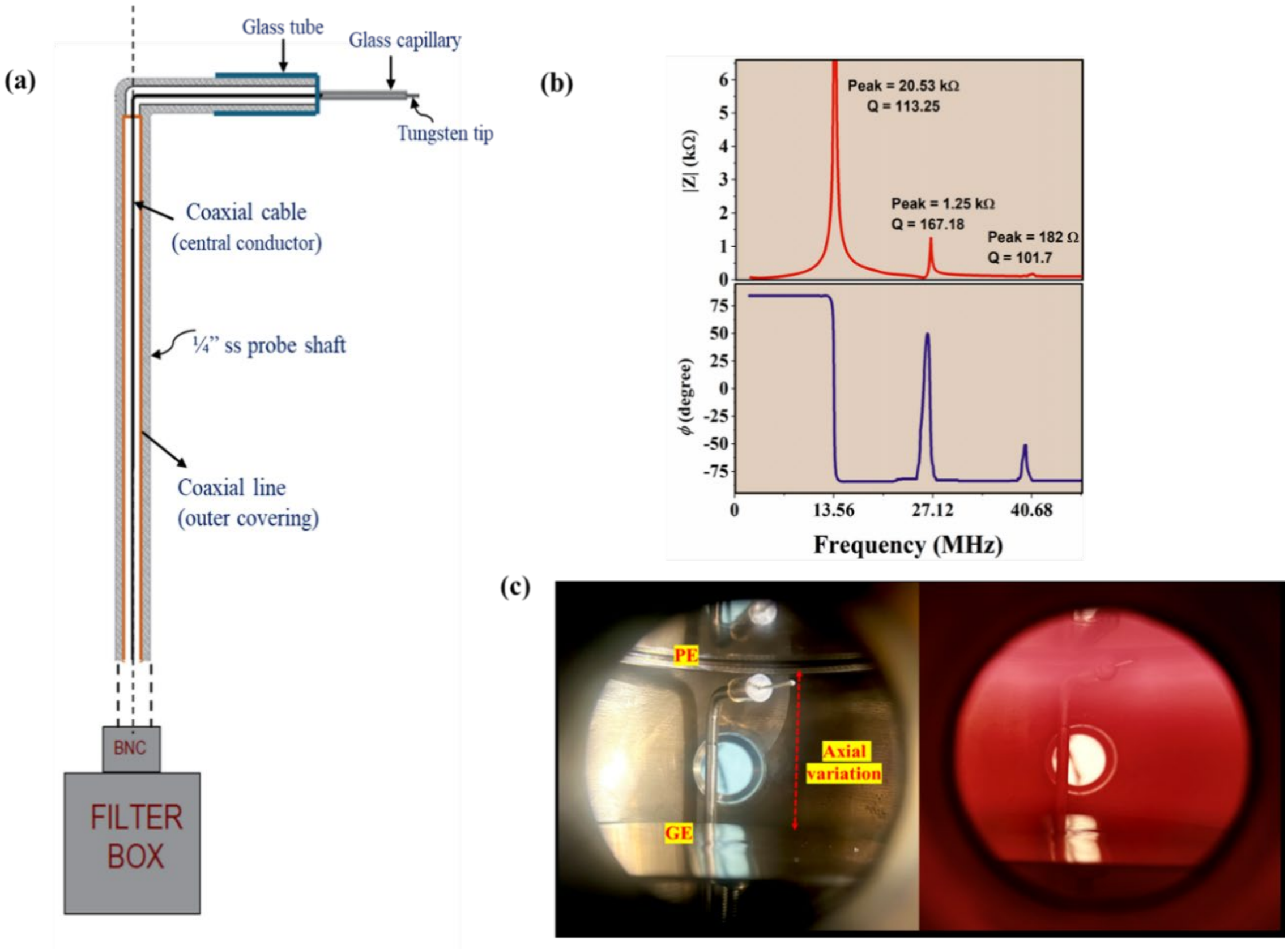

**Figure 2.** (a) Schematic of Compensated Langmuir Probe (CLP) with the tuned filter circuit. (b) Frequency response of the impedance ($Z = |Z|e^{i\Phi}$) of a three-stage tunable filter circuit connected at the end of CLP: probe impedance magnitude (top part) and phase (bottom part) (c) Photograph of plasma with installed CLP observed from side view port (LHS: without plasma to show CLP probe tip, RHS: CLP immersed in plasma discharge).

### 3.2.2 10x Voltage probe

This diagnostic is used to measure the negative self-bias voltage ($V_{DC}$) that develops on the PE at different pressures. As outlined in Sec. 2, $V_{DC}$ has been used for estimating the steady state sheath voltage drop, $V_s$ at the PE from Eq. (5). This is a simplified yet direct technique, where the tip of the 10x voltage probe (Lecroy PP018) is connected to the RF input line of the PE through an aluminum shielding box. The voltage signal is recorded on an oscilloscope (Lecroy HDO6104). The complete set-up for measurement is shown in figure 3(a). A typical oscilloscope trace shown in figure 3(b) gives self-bias voltage, $V_{DC}$ ~ -70V at 100 mTorr.

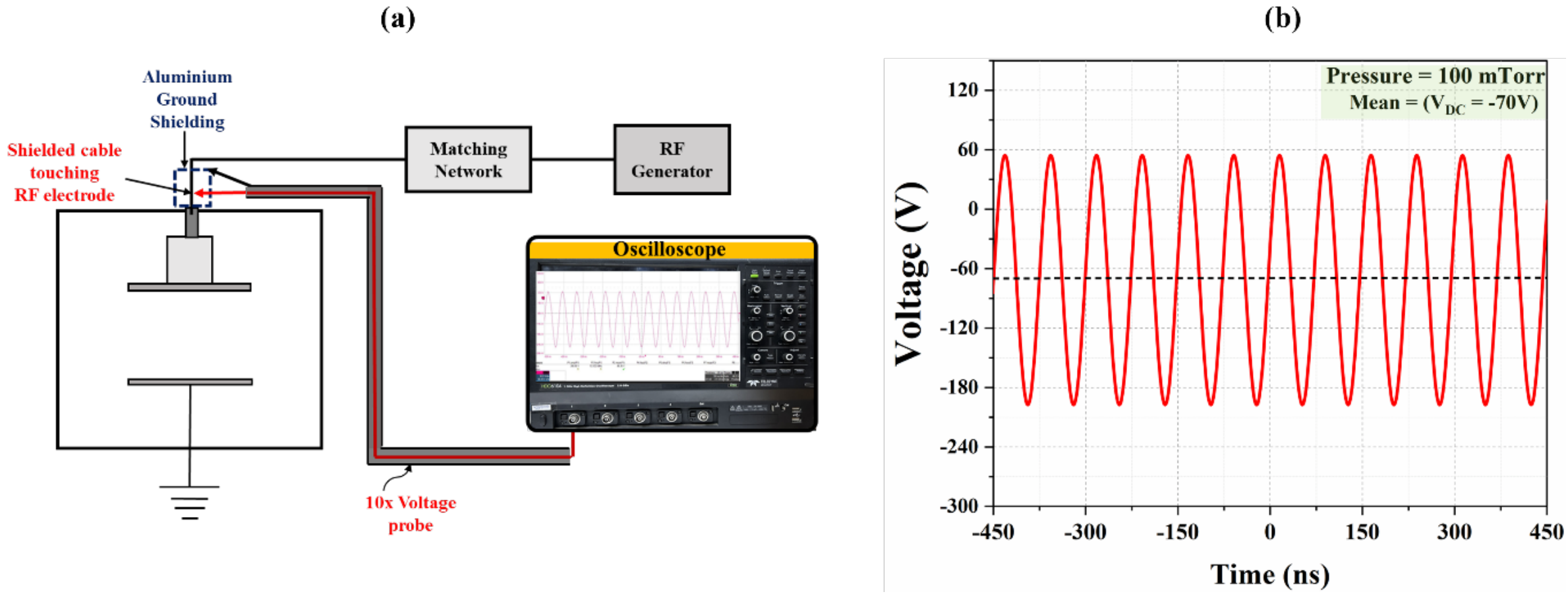

**Figure 3.** (a) Experimental setup for measurement of DC self bias voltage, $V_{DC}$ (b) Voltage trace obtained in oscilloscope (Lecroy) at 100 mTorr showing, $V_{DC} \approx -70$ V.

## 4 Experimental Results

### 4.1 Axial variation of discharge characteristics

Results of the CLP measurements are presented for a wide range of pressures (≈ 5 – 600 mTorr). It may be noted that the plasma parameters were determined from the experimentally recorded *I–V* characteristics using the standard Langmuir probe theory, as discussed in detail in Ref 23. Figure 4 displays axial variations of the argon plasma density ($n_e$), electron temperature ($T_e$), and the plasma potential ($V_p$) at ≈ 5 mTorr, ≈ 25 mTorr, ≈ 100

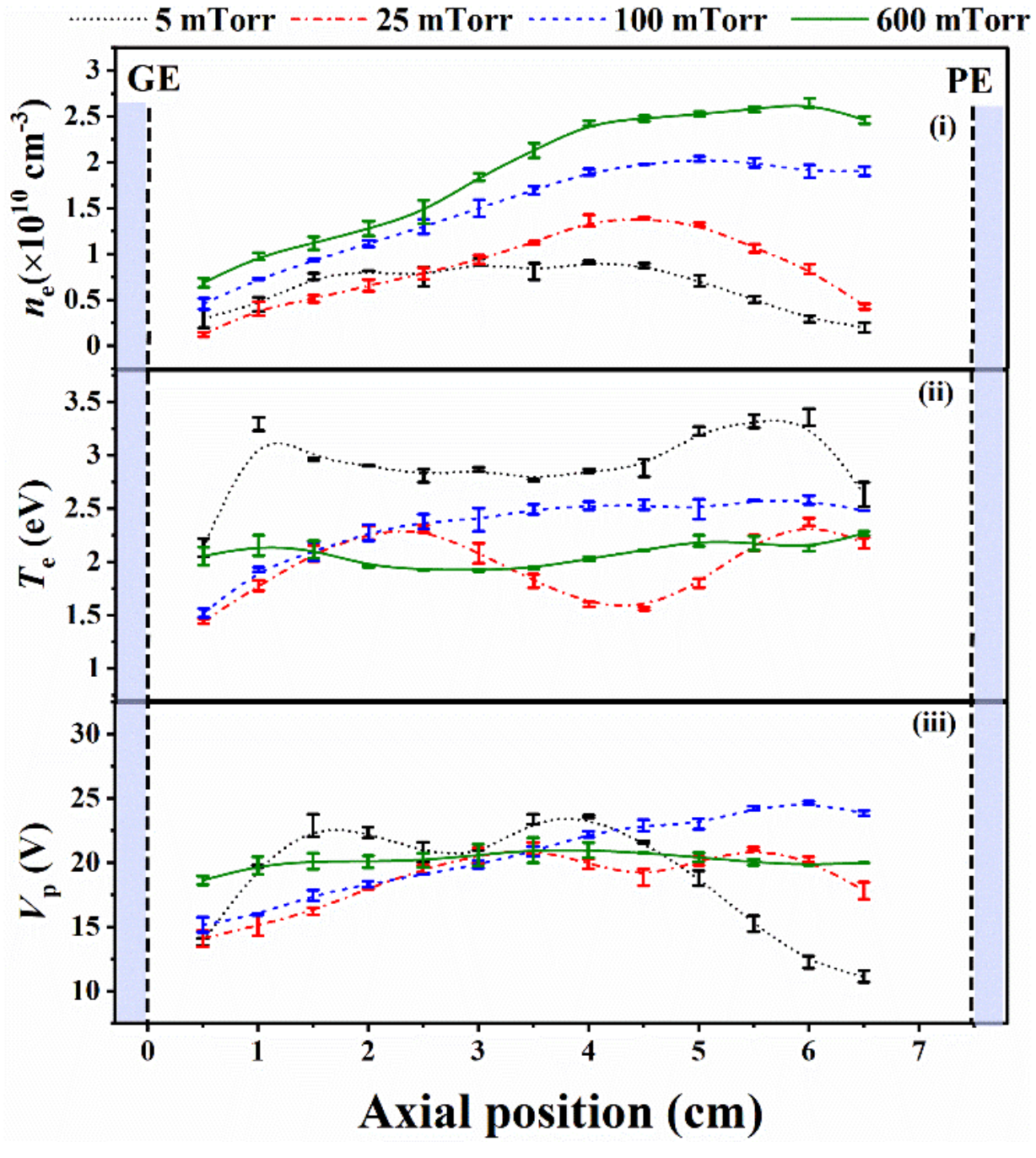

**Figure 4**. (a) Axial variation of plasma density, $n_e$ (b) Electron temperature, $T_e$ (c) Plasma potential, $V_p$ at 5 mTorr, 25 mTorr, 100 mTorr and 600 mTorr.

mTorr and ≈ 600 mTorr. It can be seen that there are significant non-uniformities in the profiles of all three plasma parameters, which tend to smoothen out with increasing pressure. It is seen from figure 4(a) that at ≈ 5 mTorr, the peak of the $n_e$ profile is almost symmetrically located (between PE and GE), with the peak shifting towards the PE with increasing pressure. For instance, at ≈ 5 mTorr $n_e$ peaks at ~ $z$ = 3.5 cm; at ≈ 25 mTorr the peak is at ~ $z$ = 4.5 cm; at ≈ 100 mTorr the peak is seen at ~ $z$ = 5.5 cm and finally at ≈ 600 mTorr the peak occurs at ~ $z$ = 6 cm. The reason for this will be examined in Section 5.

Figure 4(b) and 4(c) shows axial profiles of $T_e$ and $V_p$ at the different pressures. Both $T_e$ (≈ 1 – 3 eV) and $V_p$ (≈ 10 – 30 V) show significant axial variations at low-pressures (≈ 5 and ≈ 25 mTorr), which become increasingly uniform and flat with increasing pressure. In fact, keeping in mind that $V_p \propto T_e$, the similarity between their profiles is not surprising. It can be seen moreover, that $V_p$ being in the range, ≈ 10 – 30 V can easily confine the electrons in the bulk plasma since $T_e$ is in the range, ≈ 1 – 3 eV.

**4.2 Sheath voltage drop ($V_s$)**

The self-bias voltage $V_{DC}$ on the PE is presented in figure 5(a). As expected $V_{DC}$ (< 0) becomes less negative with increasing pressure [25-27]. It is seen to rise from ≈ – 100 V at ≈ 1 – 20 mTorr, to ≈ – 40 V at ≈ 600 mTorr. This behaviour may be attributed to a reduction of the electron–neutral collision frequency at low pressures, which enhances both electron mobility and the oscillation amplitude. Thus, with *decreasing* pressure, an increasing number of electrons are able to reach the PE, *building up the negative charge on the electrode since their path is blocked by the capacitor inside the matching network*. The net result is a boosting of the negative potential on the electrode. In steady state, $V_{DC}$ attains a value that helps check the arrival of electrons at the PE.

Figure 5(a) also shows the sheath voltage drop calculated using Eq. 5 and is given as, $V_s = V_p\,(z = 6.5) - V_{DC}$. $V_s$ increases from ≈ 40 V at ≈ 600 mTorr to about ≈ 120 V at ≈ 5 mTorr. This is acceptable because the bulk plasma potential exhibits only a modest variation with pressure (typically 10–30 V), whereas the self-bias voltage ($V_{DC}$) changes significantly.

Figure 5(b) gives axial profiles of the potential from the grounded electrode, GE to the powered electrode, PE. From figure 4 (c), $V_p$ near the GE ≈ 14 – 18 V, which renders the sheath drop at GE quite modest. On the other hand, on account of the large, negative $V_{DC}$, the sheath drop at the PE, $V_s$ is quite high. It will be seen in the next section that it is this large $V_s$ at the PE that is responsible for accelerating electrons within the sheath at the PE to sufficiently high energies for efficient ionization, even within the small length of the plasma chamber.

***Remarks:*** Before closing this section, it is fruitful to examine some relevant parameters and their variation with the pressure. Figure 6(a) gives the pressure variation of the plasma frequency, $\omega_{pe}$ and $\nu_m$, the momentum transfer frequency for elastic, *e-n* collisions. It is seen that $\omega_{pe}$ is practically constant (≈ $10^{10}$ radians/s) with the pressure, while $\nu_m$ increases almost linearly with pressure ($10^7 < \nu_m < 10^9$ s$^{-1}$) and that $\omega_{pe} >> \nu_m$ for all pressures under consideration here. The axial variations of $\nu_m$ (calculated using $\nu_m = {}^{1}/_{n_g \sigma_{el}}$, $\sigma_{el}$ taken from Ref. 22, p.73) at different pressures (shown in figure 6(b)), mirror approximately the axial profiles of $T_e$ (with the pressure in Fig. 4(b)) due to the dependence of the *e-n* collision cross sections on $T_e$. On the other hand, since the RF radian frequency $\omega \simeq 8.5\times10^7$ rad/s, $\nu_m >> \omega$ for almost the entire pressure range, there will be multiple collisions within an RF period.

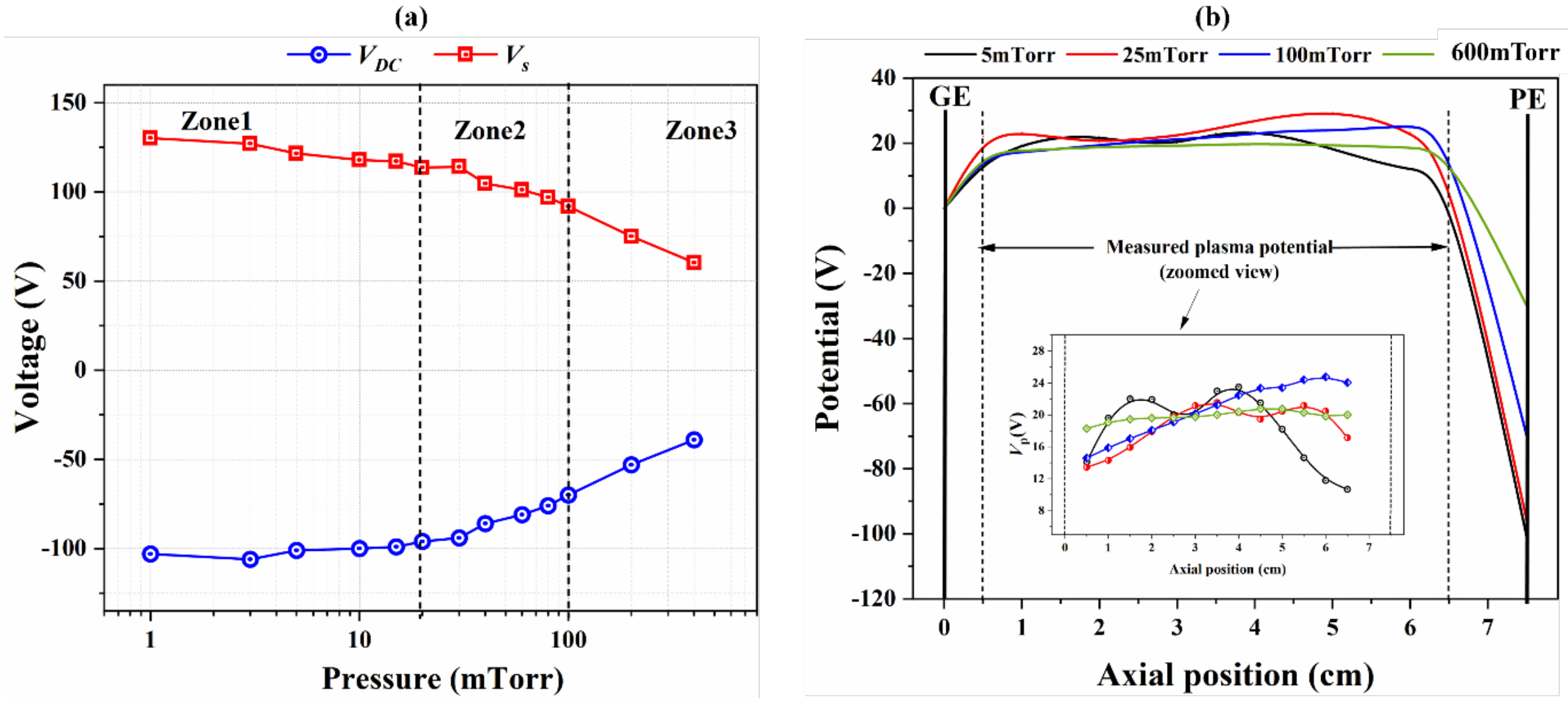

**Figure 5.** (a) Self bias voltage ($V_{DC}$) and sheath voltage drop ($V_s$) plotted with respect to pressure; (b) Potential profile between powered electrode (PE) and grounded electrode (GE) at 5mTorr, 25mTorr, 100mTorr and 600mTorr.

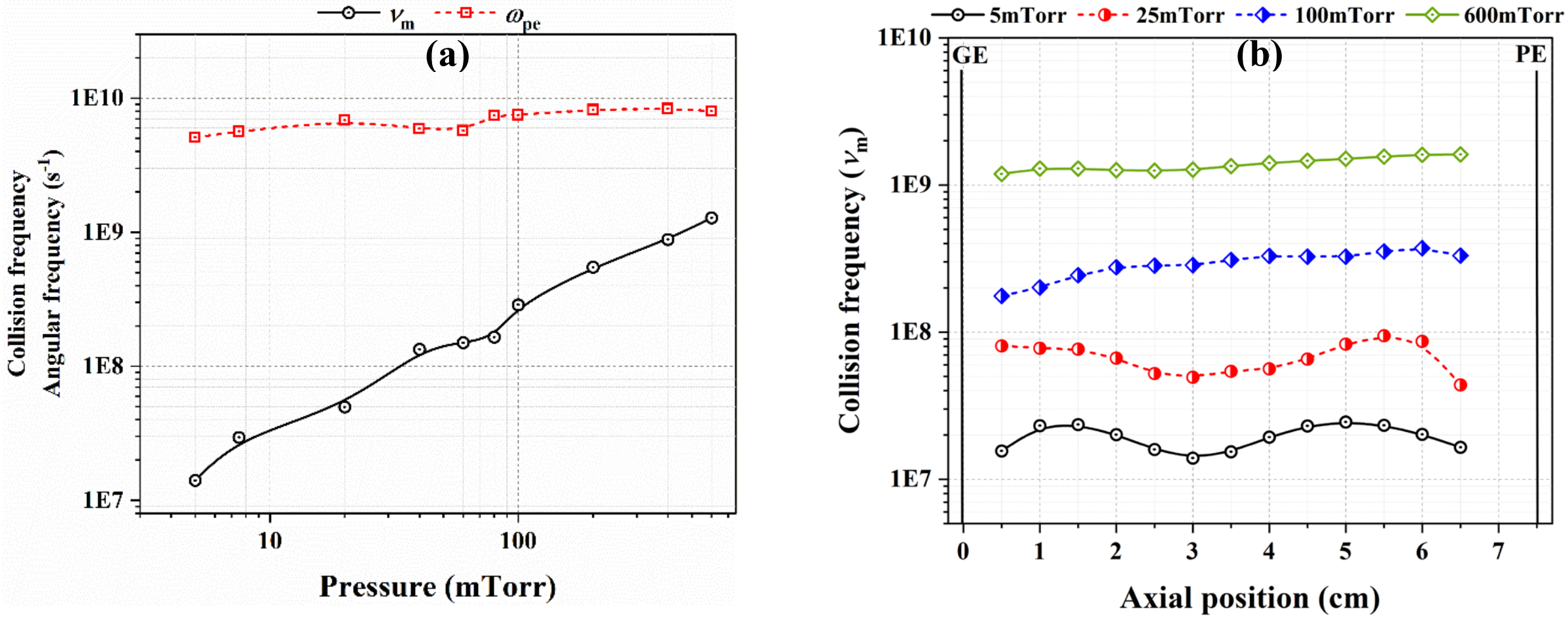

**Figure 6.** (a) Pressure variation of collision frequency ($\nu_m$) and plasma frequency ($\omega_{pe}$) (b) Axial plot of collision frequency ($\nu_m$) at 5mTorr, 25mTorr, 100mTorr and 600mTorr.

One may examine these in more detail. Noting first that $\omega_{pe} >> \omega$, it follows that the *conduction current $J_C$ in the plasma will be much larger than the displacement current $J_D$*, i.e., $J_C >> J_D$. Since in general, $\boldsymbol{\nabla}\cdot\boldsymbol{J}_{Total} = \boldsymbol{\nabla}\cdot(\boldsymbol{J}_C + \boldsymbol{J}_D) \simeq \boldsymbol{\nabla}\cdot\boldsymbol{J}_C = 0$ it follows that the *conduction current will be constant along the axis*. The conduction current in the plasma has two components: (i) an inductive part and (ii) a resistive part. However, the condition $\nu_m >> \omega$, also ensures that *the plasma conductivity will have a dominant real or resistive* part (***E*** and ***J*** almost in phase), and *not imaginary* as would happen for inductive currents (***E*** and ***J*** in quadrature phase). Thus, at the higher pressures, when the condition $\nu_m >> \omega$ is met easily, the current in the plasma would be resistive (real conductivity). Additionally, the condition, $\omega_{pe} >> \nu_m$ dictates that the *plasma conductivity will in general* *be high*, being controlled by the collision frequency (for almost constant plasma density). This also implies that ohmic power absorption will not be a dominant power absorption mechanism in CCPs. This will be examined in more detail in the next section.

One may now see how the current remains constant along the chamber. The electric field adjusts continuously along the chamber, being high where the conductivity is low and vice versa.

## 5 Discussion of Plasma Generation Mechanism

### 5.1 Ohmic Power Absorption by Electrons in Bulk Plasma

To resolve the issue of plasma production in the present experiments, it is important to examine first the Ohmic power $(P_{ohmic})$ absorbed by the bulk plasma electrons in the CCP at different pressures. To determine $P_{ohmic}$ from Eq. (1), one needs to find the current density $J$ through the plasma, which in turn is calculated by applying Eqs. (3), (4) and (5) in *reverse* sequence. Apart from the plasma parameters from which the plasma dielectric function is found, the other crucial inputs from the experiments are $V_{\mathrm{DC}}$ and $V_{ps}$ (plasma potential at steady state sheath edge at PE) from which $V_s$ is calculated (as shown in the last section).

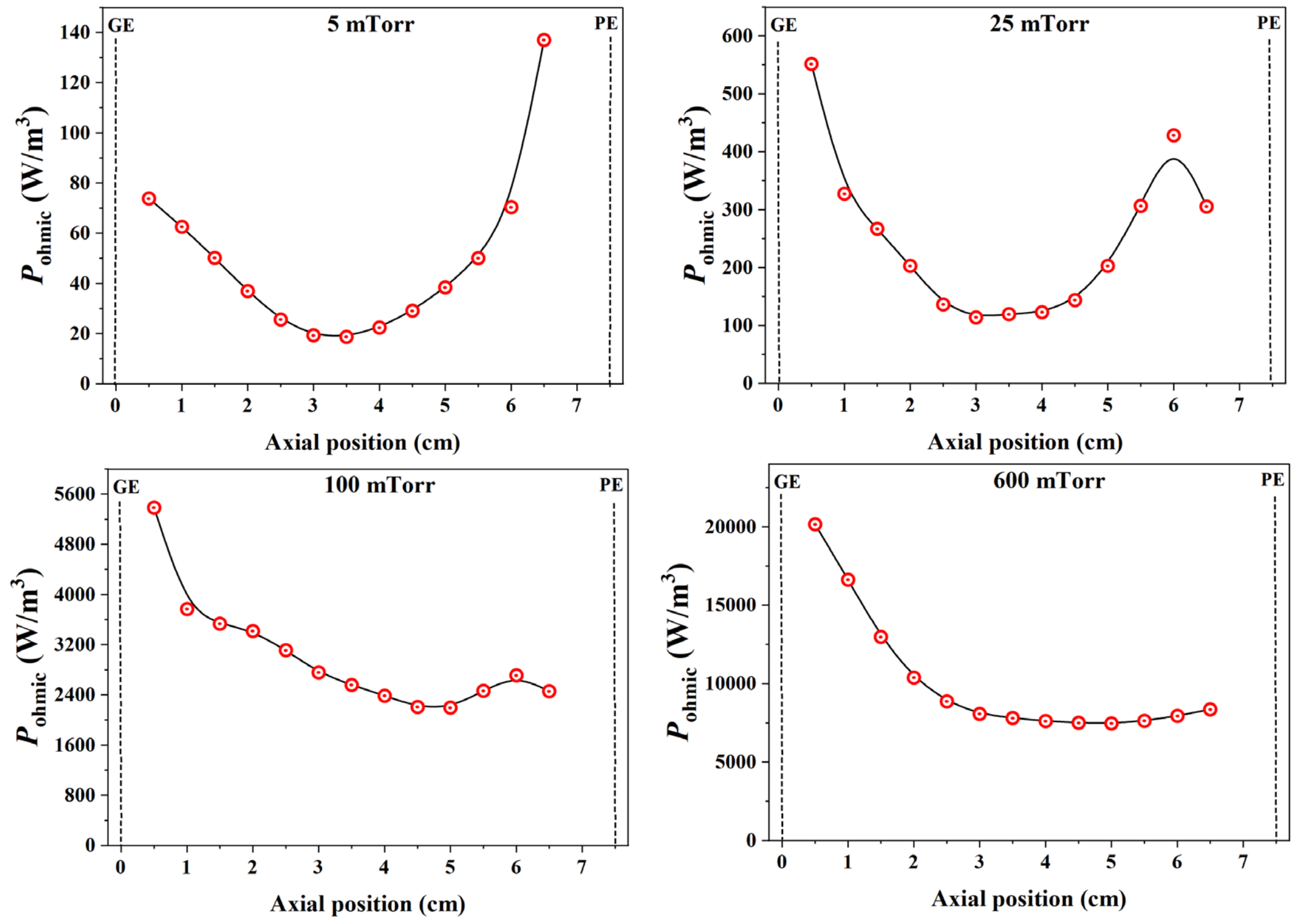

**Figure 7.** Plot showing axial profiles of local ohmic power deposition ($P_{ohmic}$) compared at (a) 5 mTorr (b) 25mTorr (c) 100 mTorr (d) 600 mTorr.

Figure 7 displays plots of $P_{ohmic}$ (W/m$^3$) at four operating pressures (≈ 5 mTorr, ≈ 25 mTorr, ≈ 100 mTorr and ≈ 600 mTorr). At low pressures (≈ 5 mTorr) one obtains a deeply concave profile with very modest values of $P_{\mathrm{ohmic}}$ (≈ 20 W / m$^3$ at the center; ≈ 80 W / m$^3$ at GE and ≈ 140 W / m$^3$ at PE). With increasing pressure, there is a rapid, overall increase in $P_{\mathrm{ohmic}}$, although the profile reverses quickly with $P_{ohmic}$ at GE increasing, until at ≈ 600 mTorr, $P_{\mathrm{ohmic}}$ ≈ 20,000 W / m$^3$ at GE and, ≈ 7,500 W / m$^3$ at PE. Comparing these plots with the density profiles in figure 4(a) shows clearly that $P_{ohmic}$ peaks in regions of low plasma density (or low conductivity) and is lowest in regions of high plasma density (or high conductivity). One can also consider non-local ionization wherein, electrons heated at a particular location can ionize at a remote location. Examining the separation between the location of peak absorption (GE or PE) and the corresponding density peak at each pressure, one notes that the separation increases with pressure (see Table 1). Regarding this separation between the location of the peak absorption and the density peak as a measure of the mean free path at each pressure, one sees that the ionization mean free path increases with gas pressure, which is not possible. Hence, one may rule out non-local ionization due to electrons heated by the ohmic mechanism. Thus, the ohmic absorption profiles establish clearly that in the present experiments *ohmic power absorption is not responsible for plasma generation.* In fact, the profiles of $P_{ohmic}$ are those that arise when an RF current flows through a *preexisting plasma medium,* and not one created by it. It follows therefore that plasma formation in the CCP must be due to another mechanism. This will be investigated in later sections.

**Table 1**. Separation between location of peak ohmic absorption and density peak at different pressures

| Pressure (mTorr) | Peak ohmic heating (from PE) | Peak plasma density (from PE) | Separation |
|---|---|---|---|
| 5 mTorr | 1 cm | >3 cm | 2 cm |
| 25 mTorr | 7 cm | 3 cm | 4 cm |
| 100 mTorr | 7 cm | 2 cm | 5 cm |
| 600 mTorr | 7 cm | 1.25 cm | ≈ 6 cm |

### 5.2 Estimation of Ionization Mean Free Paths from Probability Arguments and Density Profiles

Consider a beam of electrons passing through a field of target particles (neutral atoms). It is well known that the random event of a collision obeys the *exponential probability distribution*. Thus, the probability that an electron entering the target field at $w = 0$, *would not suffer any collision and still be in the beam at w* is $\propto \exp[-\frac{w}{\lambda}]$, where $\lambda$ is the mean free path. It follows therefore, that the probability that an electron would be *absent* at $w$, i.e., *it must have suffered a collision within distance w* is $\propto [1 - \exp\left(-\frac{w}{\lambda}\right)]$.

Applying the above argument to *ionizing* collisions with mean free path $\lambda_{iz}$, one notes that the *probability of ionization increases with w* as $\left[1 - \exp\left(-\frac{w}{\lambda_{iz}}\right)\right]$. On the other hand, *the number of electrons available for ionization depletes with w* as $\exp\left(-\frac{w}{\lambda_{iz}}\right)$. One may therefore assert that the plasma density profile will follow or track *the product of these two probabilities, i.e.,* $\sim \left[1 - \exp\left(-\frac{w}{\lambda_{iz}}\right)\right] \times \exp\left(-\frac{w}{\lambda_{iz}}\right)$. Thus, $n_e(w)$ may be expressed as,

$$n_e(w) = C\,A(w) \qquad \ldots..(7)$$

where,

$$A(w) = \left[1 - \exp\left(-\frac{w}{\lambda_{iz}}\right)\right] \exp\left(-\frac{w}{\lambda_{iz}}\right) \quad \ldots..(8)$$

and $C$ is a constant. $A(w)$ has a maximum at,

$$w_m = \lambda_{iz} \ln 2 \quad \ldots..(9)$$

One notes that since $\lambda_{iz}$ *decreases with the pressure*, so would the peak position of plasma density ($w_m$ from Eq. (9)), which would *push the density peak closer to the point of entry of the electrons or their source*, *with increasing pressure*. From Fig.4(a) one sees that *the maxima of the density profiles shift towards the PE with increasing pressure*, *suggesting that the sheath at the PE is the source of electrons responsible for ionization and plasma production in the CCP*. In light of these arguments, the origin of $w$ in Eqs. (7) – (9) was placed at the PE.

In Eq. (9), $\lambda_{iz}$ is an unknown and needs to be found from the density profiles in figure 4(a). The normalized theoretical profiles of the density in figure 8 were plotted using Eq. (7) by tweaking the value of $\lambda_{iz}$ at each pressure to match approximately the normalized experimental profiles of the density from figure 4(a) (plotted with respect to $w$, measured from the PE), while trying to ensure simultaneously that the locations of the theoretical peaks in the figures also match reasonably with the values given by Eq. (9). The range of $\lambda_{iz}$ that gives the best possible match at each pressure (along with the corresponding range for $w_m$) is also given in Figs. 8. As can be seen, reasonable values of $\lambda_{iz}$ are obtained. It is seen that the ranges of $\lambda_{iz}$ vary from (≈ 5 – 6) cm at 5 mTorr to about (≈ 1.4 – 1.8) cm at 600 mTorr.

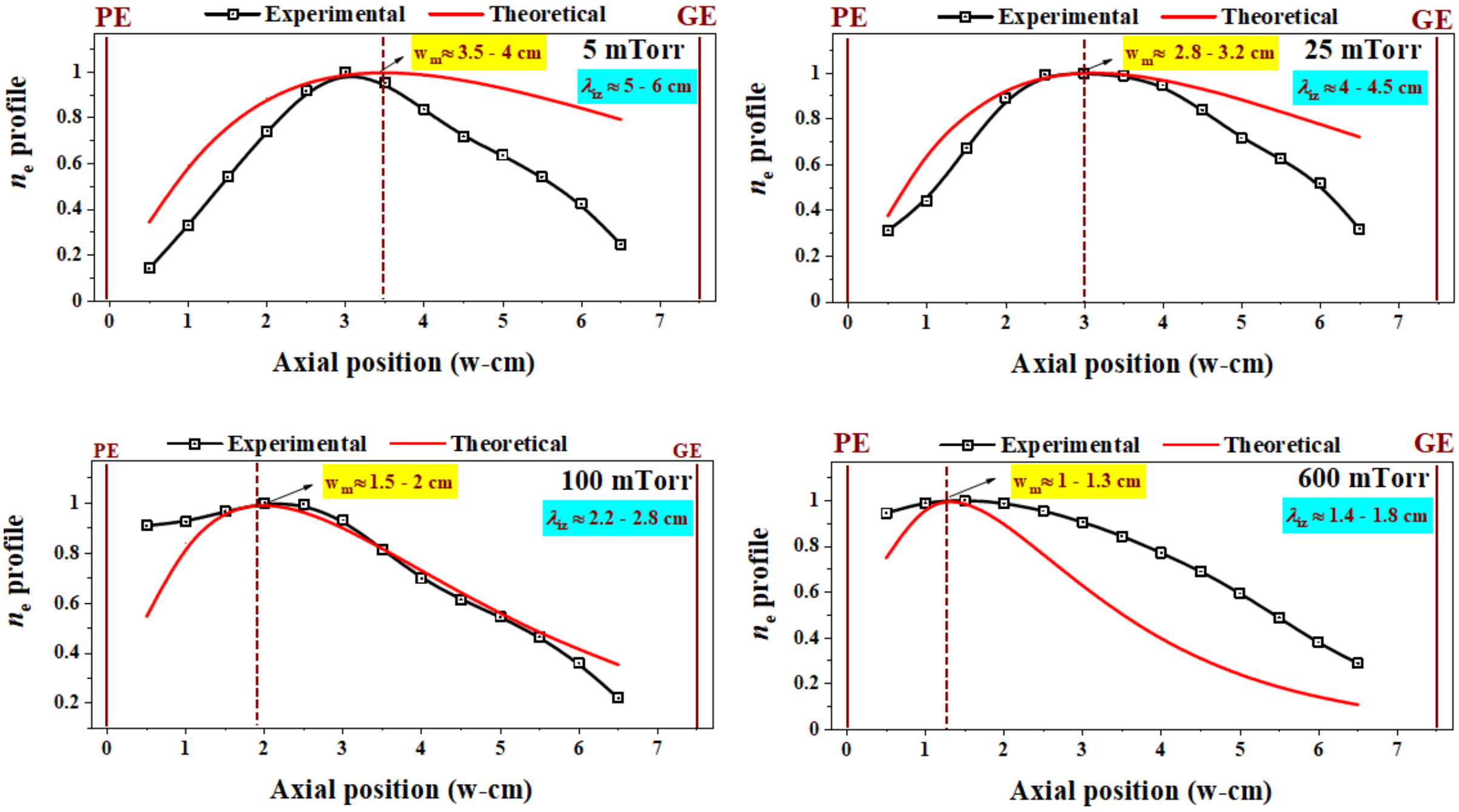

**Figure 8.** Matching the normalized experimental plasma density profiles (black) in Fig. 4(a), plotted with respect to $w$, the distance measured from the PE with the normalized theoretical profiles (red) determined from Eq. (7), by tuning $\lambda_{iz}$ to make the two profiles overlap as closely as possible. The range of values of $\lambda_{iz}$ that give the best match are given in the respective figures along with the location of the peak $w_m$, determined from Eq. (9).

***Remarks:*** It must be kept in mind that the theoretical profiles derived from Eq. (7) are *based on ionization alone and do not take into account profile modification due to particle transport* (because of density, potential gradients, etc.). The latter result in *distinctly different profiles*, particularly in steady state. This is the key reason for the difference between the experimental and theoretical profiles in figure 8.

**5.3 Estimation of Ionization Mean Free Paths from Electron Acceleration in the Sheath at the PE**

In the last section, $\lambda_{iz}$ was estimated at different pressures from the experimental plasma density data. The data also gives indication that ionization was due to electrons that acquired their energy within the sheath at the PE. This section attempts to calculate $\lambda_{iz}$ from an electron acceleration model inside the sheath at the PE, for comparison with the $\lambda_{iz}$ values in Fig. 8. The specific role of the RF in this scenario will be examined in Sec. 6.

***The Model:*** Electrons within the sheath at the PE feel strong acceleration in the steady state, DC electric field $E$ produced by the sheath voltage drop, $V_s$, shown in Fig. 5 (a). The model assumes that in steady state, electrons starting from the PE ($w$ = 0) with zero velocity ($V_e \approx 0$) are accelerated by $E$ along the sheath at the PE into the plasma. The sheath width ($s$), is given by Eq. (6). In steady state, the velocity $V_e$ obeys $\frac{\partial V_e}{\partial t} \approx 0$, where $t$ is the time. Noting that $E$ is directed from the plasma towards PE, the flow equation of motion for the electrons reads,

$$\frac{dV_e}{dt} = V_e \frac{\partial V_e}{\partial w} = \frac{eE}{m} - \nu_m V_e \qquad \ldots..(10)$$

In Eq. (10), the electron pressure term has been dropped as its contribution would be too weak in comparison to the acceleration by $E$ and the density $n_e$ has cancelled out from both sides. Also, $m$ and $e$ are the electron mass and charge and $\nu_m$ is the frequency for electron-neutral elastic collisions and is given as, $\nu_m = n_g \times K_{el}$, where $n_g$ is the argon gas density and $K_{el}$ the rate constant for *electron - atom* elastic collisions. For argon, $K_{el} \approx 10^{-13}$ m$^3$ /s, typically.

Eq. (10) is nonlinear and would need to be solved numerically from $w$ = 0 (the PE). To initiate the solution, one would need to evaluate $\frac{\partial V_e}{\partial w}$ by dividing Eq. (10) by $V_e$. However, this yields in addition to the nonlinearity, a singularity in the first term on the RHS when the initial condition, $V_e \approx 0$ is invoked at $w = 0$. These issues can be circumvented by switching from the Eulerian framework to the Lagrangian framework and solving (10) as a *linear ordinary differential equation in t*, written as.

$$\frac{dV_e}{dt} = \frac{eE}{m} - \nu_m V_e \qquad \ldots..(11)$$

Using the initial condition $V_e \approx 0$ at $t = 0$, gives the following solution for Eq. (11),

$$V_e(t) = v_d \left(1 - e^{-\nu_m t}\right) \qquad \ldots..(12)$$

In Eq. (12), $\mathrm{v_d} = \frac{eE}{m\nu_\mathrm{m}}$ is the drift speed of the electrons in the electric field, *E*. In this Lagrangian framework $\mathrm{V_e} = \frac{dw}{dt}$. Thus, one may integrate (12) to obtain the following solution for *w*(t):

$$w(t) = \mathrm{v_d}\left[t - \left(\frac{1}{\nu_\mathrm{m}}\right)\left(1 - e^{-\nu_\mathrm{m} t}\right)\right] \qquad \text{.....(13)}$$

One may eliminate the time *t* from Eq. (13) in favour of $\mathrm{V_e}$. Solving for *t* from Eq. (12) gives

$$t = \left(\frac{1}{\nu_\mathrm{m}}\right)\ln\left[\frac{\mathrm{v_d}}{\mathrm{v_d} - \mathrm{V_e}}\right] \qquad \text{.....(14)}$$

Substituting (14) into (13) obtains,

$$w\,(\mathrm{V_e}) = \left(\frac{\mathrm{v_d}}{\nu_\mathrm{m}}\right)\ln\left[\frac{\mathrm{v_d}}{\mathrm{v_d} - \mathrm{V_e}}\right] - \frac{\mathrm{V_e}}{\nu_\mathrm{m}} \qquad \text{.....(15)}$$

As stated earlier, *w* is the distance along the sheath measured from the PE. One needs to determine the velocity $\mathrm{V_{e,s}}$ of the electrons as they exit the sheath at $w \approx s$. Knowing *s* from Eq. (6) and the experimental data, it is a simple matter to solve (15) to find $\mathrm{V_{e,s}}$ at the sheath edge, $w = s$.

The sheath width and the exit velocity at the sheath edge are different for each pressure. On the other hand, to ensure that all the $\lambda_{iz}$ are referred to a common origin (the PE), *one needs to ensure that all the electrons start from a common origin (the PE* at $w = 0$) *with identical velocity, which in the present case may be taken to be the average velocity* $\mathrm{V_{e,av}}$ *of the electrons inside the sheath*. Differentiating (15) allows one to express *dw* in terms of $\mathrm{V_e}$ and $d\mathrm{V_e}$ and write the average velocity of the electrons $\mathrm{V_{e,av}}$ in the sheath, as

$$\mathrm{V_{e,av}} = \left(\frac{1}{s}\right)\int_0^s \mathrm{V_e}\, dw = \left[\frac{1}{\mathrm{v_d}\nu_\mathrm{m}}\right]\left(\frac{1}{s}\right)\int_0^{\mathrm{V_{e,s}}} \frac{\mathrm{V_e^2}}{\left[1 - \frac{\mathrm{V_e}}{\mathrm{v_d}}\right]}\, d\mathrm{v}$$

Evaluating the integral above using formula **2.111(6)** of Ref. [Gradshteyn and Ryzhik, p.69] [28] gives,

$$\mathrm{V_{e,av}} = \left[-\frac{\mathrm{v}_d^2}{s\,\nu_\mathrm{m}}\right]\left[\ln\left(1 - \frac{\mathrm{V_{e,s}}}{\mathrm{v}_d}\right) + \frac{\mathrm{V_{e,s}}}{\mathrm{v}_d}\left(1 + \frac{\mathrm{V_{e,s}^2}}{2\,\mathrm{v}_d}\right)\right] \qquad \text{.....(16)}$$

One defines $E_{\mathrm{e,av}}\,(\mathrm{eV}) = \frac{1}{2e}\, m\, V_{\mathrm{e,av}}^2$ as the energy of electrons with velocity, $\mathrm{V_e} \approx \mathrm{V_{e,av}}$. Since $E_{\mathrm{e,av}}$ is large, one may assume the rate constant for ionization $K_{\mathrm{iz}}$ to be sharply peaked about the velocity $\mathrm{V_{e,av}}$ to write,

$$K_{\mathrm{iz}} = \langle v\, \sigma \rangle_v \simeq \mathrm{V_{e,av}}\, \sigma_{\mathrm{iz}}\left(E_{\mathrm{e,av}}\right) \qquad \text{.....(17)}$$

and,

$$\lambda_{\text{iz}} = \frac{\text{V}_{\text{e,av}}}{\nu_{\text{iz}}} = \frac{\text{V}_{\text{e,av}}}{n_{\text{g}} K_{\text{iz}}} = \frac{1}{n_{\text{g}} \sigma_{\text{iz}}(E_{\text{e,av}})} \quad \text{.....(18)}$$

**Table 2.** Comparison of $\lambda_{iz}$ from the sheath model and probability arguments at different pressures.

| **Pressure (mTorr)** | **Sheath Analysis** | | **Density Profiles / Probability Argument** | |
|---|---|---|---|---|
| | **$V_{e,av}$ (m/s) [$E_{e,av}$ (eV)]** | **Ionization mean free path ($\lambda_{iz}$)** | **Ionization mean free path ($\lambda_{iz}$)** | **Location of peak density ($w_m$)** |
| 5 mTorr | $5.5 \times 10^6$ [86.32] | 15 – 16 cm | 5 – 6 cm | 3.5 – 4 cm |
| 25 mTorr | $4.34 \times 10^6$ [53.79] | 3.5 – 4 cm | 4 – 4.5 cm | 2.8 – 3.2 cm |
| 100 mTorr | $3.6 \times 10^6$ [37.05] | 2 – 2.5 cm | 2.2 – 2.8 cm | 1.5 – 2 cm |
| 600 mTorr | $2.56 \times 10^6$ [18.66] | 1 – 1.4 cm | 1.4 – 1.8 cm | 1 – 1.3 cm |

Taking $\sigma_{\text{iz}}(E)$ data for argon from Ref. [22, chapter 3] one may compute $\lambda_{\text{iz}}$ from the sheath model. The results are presented in table 2 for the different pressures. The table gives values of $\lambda_{\text{iz}}$ calculated from the sheath model and the density profiles using probability arguments [Eqs. (7) and (9)]. Also given are the corresponding values of $\text{V}_{\text{e,av}}$ [$E_{\text{e,av}}$ (eV)] and $w_{\text{m}}$.

It is seen that the $\lambda_{\text{iz}}$ values from the two models compare favourably for all pressures, except the lowest pressure, ≈ 5 mTorr, where there is a significant deviation in $\lambda_{\text{iz}}$ (differs by a factor ≈ 2.5 – 3). It turns out that the cause for the discrepancy can be traced to the failure of the above model at low pressures. This and the role of the RF at the higher pressures will be considered in the next section.

## 6 Discussion on the role of the RF at high pressures

Table 1 shows that the model presented above actually reproduces successfully the ionization mean free paths captured from the experimental density profiles in almost the entire pressure range considered in the experiments, except at the lowest pressure (≈ 5 mTorr). In what follows, an attempt will be made to see how the role played by the RF contributes to the success of the model at high pressures. Thereby, it would be possible to see why the model gives a discrepancy at low pressures.

***Role of the RF:*** To get a clearer picture regarding the role of the RF in the present scenario, it is useful to compare *the transit time* ($T_{\text{s}}$) *of the accelerated electrons across the sheath with respect to the time period* ($T_{\text{RF}}$) *of the RF.* The ratio $R$ (= $T_{\text{s}}$ / $T_{\text{RF}}$) varies in the range, $R \simeq 1.7 \times 10^{-2} - 4.6 \times 10^{-3}$ as the pressure is increased from ≃ 5 to 600 mTorr. These values of $R$ correspond to changes in phase of ≃ 6.12° at ≃ 5 mTorr, ≃ 4.12° at ≃ 25 mTorr to ≃ 1.66° at ≃ 600 mTorr. This means that at the higher pressures (≈ 25 – 600 mTorr) *during the entire flight of the electrons through the sheath there will be very little change in the RF voltage, the change decreasing with increasing pressure.* This implies that the *RF electric field seen by the electrons will be practically constant, amounting to an added DC field over and above the DC electric field due to the negative potential on the powered electrode*.

The above implies that *each electron samples an instantaneous "spot value" of the RF electric field that is completely random since it depends on the instantaneous value of the field* (or the phase of the RF). Now, *during the negative part of the cycle, electrons will draw energy from the RF* (as they do from the DC field of the PE) *while during its the positive part, the RF will draw energy from the electrons* (acting opposite to the DC field). The crucial point, however, is that the RF voltage, though symmetric with respect to the DC ground, *is not symmetric with respect to PE, since the voltage on PE adds a negative DC bias to the RF voltage. Therefore, depending on the extent of the self-bias, the RF voltage will have a negative swing for greater part of the RF period, which implies electrons sampling the field randomly are more likely draw power from the RF than pump power into it.*

All measurements in the experiments were in *steady state* over *many* RF cycles. This means that measurement of the *location* at which ionization takes place, will be averaged over many cycles of the RF, for a large number of electrons. As a result, the *spread in the location* at which ionization takes place would average to zero, yielding a fixed ionization profile at each pressure. It is noteworthy, however, although electrons sample the RF field randomly, *averaged over many cycles in steady state there will be net transfer of power from the RF to electrons*, since the negative RF voltage is asymmetric (being primarily negative for a greater part of its period).

***Populating the sheath*:** Although, the above arguments establish the role of the RF in the present context and that the electron acceleration model predicts correctly the ionization mean free paths for the higher pressures (≈ 25 – 600 mTorr), one still needs to consider as to how electrons would populate the sheath at the PE. As outlined in Ref. 22, there is a time-window during the positive swing of the RF voltage that permits the PE to become positive with respect to the plasma, producing a receding sheath. This allows the electrons to be attracted into the sheath and hence, populating it.

Another way the sheath could be populated is by *secondary emission of electrons* from the PE, when ions accelerated by the sheath voltage drop bombard the PE. For low energy ions (< 200 eV), the secondary ionisation coefficient $\gamma_{\text{sec}}$ for argon ions on stainless steel (ss) is typically ≃ 0.05 [29,30]. One may compute the ion particle flux using the Bohm flux from which one may determine the net secondary emission flux. These electrons would be accelerated by the DC potential on the PE. Normally, these secondaries would ionise the gas within the sheath itself, leading to avalanching. However, assuming no ionization inside the sheath (as is the situation in the present scenario), the secondary electron flux would be conserved. Since the velocity at the sheath edge is known one may determine the density of secondary electrons at the sheath edge. Doing this, one finds that the density of secondary electrons is about two orders smaller than the sheath edge density. Consequently, one may rule out the secondary emission process in the present context as a means to populate the sheath. Moreover, the power required for discharges to be sustained by secondary emission are considerably higher than that used here (≃ 10 W). In general, the characteristics of plasmas sustained by secondary emission are distinctly different from those observed in the present experiments.

## 7 Conclusion

This work investigates plasma production in capacitively coupled plasma (CCP) by measuring the axial profiles of the plasma parameters and the DC self-bias voltage on the powered electrode (PE) over a wide range of pressures (≈ 5 – 600 mTorr). It is shown at the outset that Ohmic power absorption by the electrons in the bulk plasma is not responsible for plasma production in the CCP. A model based on electron acceleration by the self-bias voltage in the DC sheath at the powered electrode was developed to calculate the ionization mean free paths ($\lambda_{\text{iz}}$) at the different pressures. It is shown that these compare favorably with the $\lambda_{\text{iz}}$ extracted from the plasma density axial profiles for pressures from ≈ 25 mTorr ($\nu_m \simeq \omega$) to ≈ 600 mTorr ($\nu_m >> \omega$). It is also shown that the electrons, due to their extremely short sheath transit times (relative to the RF period), sample only instantaneous "spot values" of the RF field, amounting to an added DC field over and above the DC electric field due to the negative potential of the PE. The negative bias on the electrode renders the RF swing asymmetric making it negative for most of the cycle. Since the RF accelerates electrons when it is negative and takes energy from them when it is positive, net power is transferred to the electrons during the course of steady-state measurements, *averaged over many RF cycles*.

It may be noted that the ionisation mean free paths at the lowest pressure (≈ 5 mTorr) determined from the model show significant deviation from those determined from the plasma density profiles showing that the mechanism proposed in the model does not hold at this pressure. Since, $\nu_m < \omega$, at this pressure, it is most likely that the stochastic heating mechanism is at work here.

### Acknowledgment

The experimental system was partially supported by Department of Atomic Energy, India (BRNS) [BRNS Ref: 2012/34/4/BRNS/125, IITD ref: RP02601] and the above work was also partially supported from Core Research Grant (CRG), Science & Engineering Research Board (SERB), DST [SERB Ref: CRG/2021/004273, IITD ref: RP04275G].